\documentclass[sigconf]{acmart}

\usepackage{pifont}
\usepackage{amssymb}
\usepackage{balance}
\usepackage{graphicx}
\usepackage{dblfloatfix}
\usepackage{url}
\usepackage{multirow}
\usepackage{multicol}
\usepackage{subcaption}
\usepackage{color}
\usepackage{amsmath}
\usepackage{mathtools}
\usepackage{hyperref}
\usepackage{enumitem}
\hypersetup{
  colorlinks = true, 
  urlcolor   = blue, 
  linkcolor  = blue, 
  citecolor  = red   
}
\def\UrlBreaks{\do\/\do-}
\newcommand{\cmark}{\ding{51}}
\newcommand{\xmark}{\ding{55}}

\definecolor{grey}{rgb}{1,0.7,0.5}
\definecolor{purple}{rgb}{1,0,1}
\definecolor{light-blue}{rgb}{0,1,1}

\renewcommand\footnotetextcopyrightpermission[1]{} 
\begin{document}

\title{Shortcuts through Colocation Facilities}

\author{Vasileios Kotronis}
\affiliation{
  \institution{FORTH, Greece}
}
\email{vkotronis@ics.forth.gr}

\author{George Nomikos}
\affiliation{
  \institution{FORTH, Greece}
}
\email{gnomikos@ics.forth.gr}

\author{Lefteris Manassakis}
\affiliation{
  \institution{FORTH, Greece}
}
\email{leftman@ics.forth.gr}

\author{Dimitris Mavrommatis}
\affiliation{
  \institution{FORTH, Greece}
}
\email{mavromat@ics.forth.gr}

\author{Xenofontas Dimitropoulos}
\affiliation{
  \institution{FORTH, Greece}
}
\affiliation{
  \institution{University of Crete, Greece}
}
\email{fontas@ics.forth.gr}

\begin{abstract}
Network overlays, running on top of the existing Internet substrate, are of perennial value to Internet end-users in the context of, e.g., real-time applications.
Such overlays can employ traffic relays to yield path latencies lower than the direct paths, a phenomenon known as Triangle Inequality Violation (TIV). Past studies identify the opportunities of reducing latency using TIVs. However, they do not investigate the gains of strategically selecting relays in Colocation Facilities (Colos).
In this work, we answer the following questions:
(i) how Colo-hosted relays compare with other relays as well as with the direct Internet, in terms of latency (RTT) reductions; (ii) what are the best locations for placing the relays to yield these reductions.  
To this end, we conduct a large-scale one-month measurement of inter-domain paths between RIPE Atlas (RA) nodes as endpoints, located at eyeball networks. We employ as relays Planetlab nodes, other RA nodes, and machines in Colos.
We examine the RTTs of the overlay paths obtained via the selected relays, as well as the direct paths. We find that Colo-based relays perform the best and can achieve latency reductions against direct paths, ranging from a few to 100s of milliseconds, in 76\% of the total cases; $\sim$75\% (58\% of total cases) of these reductions require only 10 relays in 6 large Colos.  
\end{abstract}



\copyrightyear{2017}
\acmYear{2017}
\setcopyright{acmlicensed}
\acmConference{IMC '17}{November 1--3, 2017}{London, United Kingdom}
\acmPrice{15.00}
\acmDOI{10.1145/3131365.3131388}
\acmISBN{978-1-4503-5118-8/17/11}

\maketitle

\renewcommand{\shortauthors}{Vasileios Kotronis et al.}
\section{Introduction} \label{section:intro}

\setlist{nolistsep}

Every millisecond of Internet latency counts. A broker could lose \$4 million with every passing millisecond (ms), if their electronic trading platform lags 5 ms behind the competition~\cite{amazonlatency}.

Overlay networks have historically been used to attach desirable properties to the classic best-effort Internet, including lower latency~\cite{jiang2016via}; reliability~\cite{andersen}, security~\cite{peter}, avoidance of certain areas via ``detours''~\cite{levin2015alibi}, and higher throughput~\cite{cai2016cronets} are only some of them. Operating over a stable IP-based underlay, overlays have revolutionized the way the Internet is used in the last decades. In particular, end-users, as well as their overlay application providers, have much to gain from low-latency overlay paths for real-time applications such as online gaming~\cite{ly2010improving,ly2011irs}, VoIP~\cite{amir2006overlay,jiang2016via}, and financial transactions~\cite{laughlin2014information}. 
Such end-users typically reside in \emph{eyeball} networks, namely access ISPs at the last mile~\cite{sundaresan2016home}.  

Two important research questions that cut through most efforts studying overlay networks (see Section~\ref{section:related-work}) are the following: ``\emph{What are the best locations to place overlay TIV relays, in order to improve performance or resiliency? What are the quantified benefits of choosing these relays instead of others?}''. End-hosts in eyeball networks and dedicated servers in PlanetLab are common relay choices in real systems (such as p2p networks) and academic studies. 

In this work, focusing on the latency-wise improvement of Internet paths, we examine the increasingly popular~\cite{chatzis2015quo,giotsas2017detecting,giotsas} colocation facilities (Colos) as relay sites.  
Colos provide space, power, cooling, and physical security for the server, storage, and networking equipment of colocated companies, connecting them to cloud /content providers, transit networks and eyeball ISPs, as-a-service, in multiple locations worldwide~\cite{equinix}.
Thus, they host layer-2/3 interconnections (such as IXPs~\cite{giotsas}), ranging from private to public multilateral peering setups among different ISPs. The pervasiveness of Colos on a global level has brought Internet organizations (and their users) closer to each other, driving Internet flattening~\cite{dhamdhere2010internet}.
In addition, the ecosystem of colocated companies has evolved in recent years to include many small and medium cloud providers that house their equipment (compute servers, storage, etc.) in the Colo. This change allowed for the first time in Internet's history third parties, such as end-users or application service providers, to easily rent Virtual Machines (VMs) in the largest Colos using the services of cloud providers~\cite{digital-ocean}.

Towards understanding the implications of this change for delay-sensitive overlay services, such as Skype and Hola, we investigate how the performance of Colo relays compares with other types of relays.
Colos might be considered quite promising candidate TIV relays due to their core networking location. However, it is not straightforward to expect that Colos always perform better; moreover, their exact benefit should be properly quantified. To investigate this further, we choose endpoints within eyeball networks, utilizing 3 different types of relays (see Section~\ref{section:measurement-methodology}): Colo relays (interfaces located in facilities~\cite{giotsas}), PlanetLab nodes (mainly at research institutes) and RIPE Atlas nodes (at eyeball and other networks). 
We simulate stitching of paths between endpoints through these relays, based on RTT measurements, to form single-relay TIV paths~\cite{han}. We compare the formed overlay paths with each other as well as with the direct BGP-derived paths, in terms of latency. 

Based on a large-scale, 1-month measurement campaign, and after verifying eyeball networks (see Section~\ref{subsection:endpoints-selection}) and Colo locations (see Section~\ref{section:colo-selection}) for accuracy, we identify that the best locations for placing relays are actually the Colos (see Section~\ref{section:measurement-results-analysis}). To the best of our knowledge, we are the first to study the impact of Colos w.r.t.~latency at a large scale.
We observe that Colo-relayed paths can yield median latency improvements of $>$10ms (with a 6\% of the cases gaining $>$100ms) vs.~the direct paths, and in contrast to the other relays (58\% for RIPE Atlas, 43\% for PlanetLab), improve 76\% of the total studied cases. Interestingly, a relatively small number of Colos ($\sim$6) is required to achieve most of these gains ($\sim$75\% of improved cases, 58\% of total), while the rest of the studied overlays need to employ one order of magnitude more relays to reach their respective top performance. Moreover, relaying through different countries (compared to the endpoints) helps reduce latency, probably due to the forced discovery of alternate, non-inflated~\cite{spring2003causes} BGP paths. We further show that our insights are consistent over time. 

\section{Measurement Methodology} \label{section:measurement-methodology}

The objective of our measurement methodology is twofold:
(i) to study the path latency obtained by employing relays as intermediate nodes of overlay inter-domain TIV paths, and (ii) to assess the benefits of selecting vantage points at large Colos as Internet relays, vs.~relays placed at the Internet's eyeball (and other) networks. To this end, we employ a real-world, Internet-wide testbed, comprising:
\begin{itemize}[noitemsep]
  \item \textbf{Endpoint nodes}: a set of globally distributed nodes, acting as source \emph{src} and destination \emph{dst} nodes of Internet inter-domain paths.
  \item \textbf{Overlay relays}: a set of relay nodes that are employed as intermediate hops within an inter-domain path between a \emph{src} and a \emph{dst}.
  \item \textbf{Inter-domain overlay links}: logical links that connect pairs of nodes (endpoints and/or relays) over the physical network of one or more intermediary ISPs. The underlying paths are typically derived by BGP.
\end{itemize}

We next describe how we select \emph{src} and \emph{dst} endpoints in eyeball networks (see Section~\ref{subsection:endpoints-selection}), relays at colocation facilities (see Section~\ref{section:colo-selection}) and relays at other locations (see Section~\ref{section:other-selection}). We further explain a strategy to limit the number of candidate relays based on their relative position to the endpoints in Section~\ref{subsection:limiting-relays}. Finally,
we unfold our complete measurement framework, including the applied setup and workflow in Section~\ref{subsection:measurement-framework}.

\subsection{Selection of Endpoints at Eyeballs} \label{subsection:endpoints-selection}

To perform our measurement campaign, first we select pairs of endpoints (both in different countries), which communicate either directly or via relays.
For this purpose, we use RIPE Atlas (RA) \cite{ripe-atlas} nodes (i.e., probes and anchors), a globally distributed measurement infrastructure consisting of end-host devices, capable of conducting different types of data-plane measurements.

\begin{figure}[t!]
    \centering
    \includegraphics[width=1.0\columnwidth]{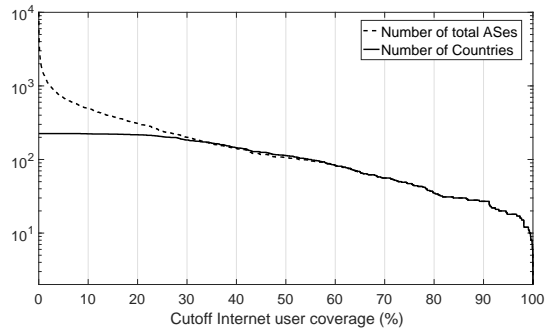}
    \vspace{-5mm}
    \caption{Number of covered ASes/countries (log-scale) worldwide \textit{vs.}~the cutoff Internet user coverage (coverage for each AS in its respective country of operation).}
    \vspace{-5mm}
    \label{fig:cutoff_coverage}
\end{figure}

Since end-users primarily reside in eyeball networks, we want to select RA nodes located within eyeballs, hence close to the end-user. Our aim is not to exhaustively cover all eyeballs, but to find a set of ASes, with sufficient country/AS-level diversity, qualifying as such.
To find these ASes, we utilize the results of the IPv6 measurement campaign by APNIC~\cite{apnic-ipv6}. 
The dataset contains 19857 ASes from 225 countries.
Besides IPv6 adoption and statistics related specifically to IPv6 users, APNIC estimates Internet user population percentages (i.e., user coverage) per AS per country for both IPv4 and IPv6 combined.
These percentages drive the eyeball selection process as follows.
The measured ASes face Internet users---browsing the web---but in order to characterize them as actual ``eyeball'' ISPs (and not e.g., enterprise networks), we also require a sufficient percentage of user population per country served by the ASes (i.e., a ``cutoff'' coverage).
Note that a large cutoff coverage may support the eyeball characterization, but can exclude countries with fragmented eyeball ISP ecosystems (e.g., the US). In Fig.~\ref{fig:cutoff_coverage}, we show the number of covered countries and ASes worldwide versus the cutoff threshold. 
If there is an AS present in a certain country, with a given coverage level, then the country is considered covered at this level. Almost all countries (223/225) as reported in APNIC's dataset, host at least one AS serving more than 10\% of the respective country's user population; 494 ASes satisfy this threshold, offering relative diversity.
Above $\sim$30\%, the 2 lines (see Fig.~\ref{fig:cutoff_coverage}) converge, indicating that only 1 AS per country is present, yielding a low AS-level diversity.
We validate if the 10\% threshold is an appropriate lower bound for ASes to be considered as eyeballs (within their respective countries).
We successfully verify all \emph{494 ASes} by \emph{manually} examining their official websites, and discovering Internet services provided to end-users (e.g., last-mile access).

We then select as endpoints RA nodes which belong to the verified $(ASN,CC)$ tuples ($CC$ = country code of AS; a single eyeball AS may be present in multiple countries). We consider only RA nodes that are: 
(i) running the latest RA firmware version
to minimize interference across measurements, affecting older versions~\cite{holterbach}, (ii) publicly available, (iii) connected and pingable,
(iv) tagged with their geolocation coordinates, and (v) stable, connectivity-wise, during the last 30 days. This filtering yields $\sim$1190 probes, associated with 141 ASes at 82 countries (where RA is present).
For each measurement round (cf. Section~\ref{subsection:measurement-framework}) we perform sampling on this population by selecting randomly: (i) one eyeball AS per country, and (ii) one node from this AS as RA endpoint ($RAE$).
This 2-step sampling limits the number of endpoints per round to a reasonable number of 82 $RAE$s on average, while preserving endpoint diversity and not incurring the bias of eyeballs with dense probe deployments. 
While this strategy may represent smaller vs.~larger countries with the same number of $RAE$s, our goal is to achieve country-level instead of complete geographical/population diversity.

\subsection{Selection of Relays at Colos} \label{section:colo-selection}

For this study, our objective is to use relays located at multiple Colos around the globe. 
We require pingable IPs, located at Colos, to use as ping targets. Root or user access e.g., to VMs hosted in colocated clouds is not necessary for the purposes of this latency-oriented study; sizable pools of colocated IPs, belonging either to routers or servers, remaining stable over time, suffice\footnote{However, relay implementations could be hosted on colocated clouds (cf. Table~\ref{tab:top-fac}).}.

To generate such a pool of IPs, we use the publicly available dataset produced by Giotsas \emph{et al.}~\cite{giotsas-dataset,giotsas} in 2015.
The authors identified facility crossings by applying their constrained facility search algorithm on extensive traceroute measurements, achieving an accuracy of over 90\%, outperforming  heuristics  based  on  naming  schemes  and IP geolocation. The data provide IP addresses that belong to facility members and are present at a candidate set of facilities, together with their respective ASN and neighboring IXPs. However, due to the age of the dataset, 
we need to exclude stale information by applying, in-order, the following filters.

\textbf{Single-facility \& active PeeringDB presence.} Preserve only the IP addresses for which the set of candidate facilities contains exactly one facility
that is still present in PeeringDB~\cite{peeringdb} today\footnote{
The facility mapping algorithm of~\cite{giotsas} may yield more than one facility for a single IP, due to inability to converge; therefore we select only active single-facility mappings to eliminate the possibility of using the wrong facility.}.
\textbf{1008} out of the initial 2675 IP addresses pass this rule.

\textbf{Pingability.} Preserve only the IP addresses that are still pingable (after a period of almost two years).  
\textbf{764} out of the previous 1008 IP addresses pass this rule.

\textbf{Same IP-ownership.} Preserve only the IP addresses whose ASN is the same as given in the initial dataset~\cite{giotsas-dataset}, since the IP-to-ASN mapping needs to be consistent.
We also check that this IP is not simultaneously advertised by multiple ASes (MOAS) to increase confidence in the dataset.
To verify this, we use CAIDA's AS-to-Prefix dataset~\cite{caida-as-to-prefix} to map IPv4 prefixes to ASNs.
\textbf{725} out of the previous 764 IP addresses pass this rule.

\textbf{Active Facility presence of ASN}. We preserve the IP addresses whose verified ASN owner is still present at the candidate facility, according to PeeringDB data~\cite{peeringdb}. \textbf{725} out of the previous 725 IP addresses pass this rule.

\textbf{RTT-based geolocation.}
First, we extract the facility's city from PeeringDB~\cite{peeringdb}.
Our 725 candidate IP addresses are associated with 103 facilities present at 67 cities around the globe. 
We want to ensure that these IPs are located at the respective city of the candidate facility.
Current IP-based geolocation services do not provide city-level accuracy~\cite{maxmind_acc,poese2011ip,shavitt2011geolocation}, thus, to determine each IP location we use Periscope, a tool 
that utilizes available Looking Glass servers (LGs)~\cite{giotsas2016periscope} (1818 LGs at 526 cities at the time of measurement, i.e., between 1-6 April 2017). 
For each candidate IP, and for each set of LGs residing in the same city as this IP's facility,
we measure the RTT from the LGs towards the IP address. Since Periscope currently supports only traceroute probes from LGs,
we calculate the RTT as the one yielded on the last hop to the IP.
We keep the minimum RTT for each IP as the primary indicator, to avoid RTT inflation effects affecting other LGs. We consider only IPs for which Periscope measurements are available and for which the minimum RTT does not exceed a threshold of 1ms~\cite{singla}.

The rule-checking process yields \textbf{356} IP addresses mapped to 58 facilities in 36 cities around the world (US, Europe, SE Asia and Australia).
During each measurement round we select randomly 1 to 3 IPs per facility to both cover all available facilities and account for variance within facilities, thus working with a sampled population of 129 IP addresses on average, used as Colo relays ($COR$).

\subsection{Selection of Relays at Other Locations} \label{section:other-selection}
Except for overlay relays residing at Colos, we consider relay nodes hosted at other locations as alternative Internet vantage points. To this end, we use publicly available nodes from PlanetLab \cite{chun} (see Section~\ref{section:planetlab-selection}) and RIPE Atlas \cite{ripe-atlas} (see Section~\ref{section:ripe-relay-selection}).

\subsubsection{PlanetLab Relays} \label{section:planetlab-selection}

We extract the first set of relays from PlanetLab \cite{chun}, a global research network that numbers $\sim$1.4k nodes (at 717 sites), mostly located in research and academic institutions.
Having allocated 500 nodes from 62 sites as candidate relays out of this set, we select randomly 1 to 2 nodes per site that are consistently accessible and pingable before each measurement round, thus working with an average sample of $\sim$59 PlanetLab relays ($PLR$)\footnote{
Despite the number of sampled $PLR$ being smaller than the corresponding $COR$ samples due to issues with the availability of functional PlanetLab nodes, both relay sets have geo-presence at a comparable number of sites ($\sim$60).}.

\subsubsection{RIPE Atlas Relays} \label{section:ripe-relay-selection}

We employ two independent sets of RA relays ($RAR$), the one from nodes at eyeball networks ($RAR\_eye$) and the other from nodes at networks that have not been verified as such ($RAR\_other$), potentially in core locations~\cite{ripe-probe-locations}.

\textbf{Eyeball Networks.} To generate the eyeball relay set, we follow the methodology of Section~\ref{subsection:endpoints-selection}. We then sample 82 relays (as many as the available countries) on average for each measurement round.

\textbf{Other Networks.} For the other relays, we use all the rest of the available $(ASN,CC)$ tuples. Out of $\sim$2500 remaining relays, we randomly select one relay per country, gathering 102 relays on average per measurement round.

\subsection{Choosing Feasible Relays} \label{subsection:limiting-relays}

Not all available relays are useful for a certain pair of endpoints.
Some of them, even if used under ideal conditions within a ``speed-of-light'' Internet~\cite{bozkurt2017internet}, still yield larger latency than the observed direct path.
Thus, to exclude such relays, we follow a simple approach based on the geolocation information of the involved nodes. 
Given a certain pair of endpoints ($n_1$, $n_2$), we compute the geographical distance $d(n_1,n_2)$ between them and then the propagation delay $t(n_1,n_2)=d(n_1,n_2)/(c*\frac{2}{3}$, for the speed of light in an optical fiber~\cite{singla}). If $RTT(n_1,n_2)$ is the measured RTT between the two endpoints, we keep only the \emph{feasible} relays~$f$ that satisfy:
\begin{equation*}
2*[t(n_1,f) + t(f,n_2)]\leq RTT(n_1,n_2)
\end{equation*}

\subsection{Measurement Framework} \label{subsection:measurement-framework}

We measure RTT as the metric for inter-domain path latency, using pings between the following pairs of nodes. 

\textbf{Endpoint-to-endpoint.}~Pairs of $RAE$ nodes to measure latency of direct, BGP-derived, Internet paths.

\textbf{Endpoint-to-relay.}~Pairs of $RAE$ and relay nodes to measure latency on overlay links, which may be stitched to form alternative paths from/to endpoints traversing a relay (i.e., relayed paths). Relays can be $COR$s, $PLR$s, and $RAR$s.

First, measuring the RTTs via pings between each pair of nodes for both directions, we observed that the direction of the ping does not affect the RTT. For example, for
$\sim$80\% of the $RAE2RAE$ cases, the difference between initiating the ping from one node instead of its counterpart does not exceed 5\%, while it is averaged out to $\sim$0\% due to our randomized pair selection strategy.

We base our measurements on the following principles: 
(i) work under the RA measurement constraints~\cite{ripe-atlas-constraints},
(ii) amortize timing differences between pseudo-parallel measurements, due to self or external interference~\cite{holterbach} and lack of synchrony, via randomized setups, use of
median\footnote{Since RA is a best-effort measurement infrastructure w.r.t.~synchrony and concurrency~\cite{holterbach}, we use batches of measurements and search for representative pairwise RTTs. To avoid distorting the results with heavy outliers (which exist), we use the median, instead of the average, as a robust metric to represent each batch.} values, and long repeated experiments.
We thus schedule measurements between endpoints as well as endpoints and feasible relays, via the RIPE Atlas API~\cite{ripe-atlas-api},
repeating a 4-step workflow (\emph{round}) every 12 hours (20 April - 17 May 2017) to account for diurnal patterns. 
The basic measurement pattern lasts 30 minutes; this window was chosen large enough to account for RTT variability, and small enough to encapsulate a sufficiently correlated batch of measurements. During this window, pings in 5-minute intervals are sent between each pair of nodes, generating an adequate number of measurements (6 per pair) to properly evaluate the associated median RTTs. The workflow steps are:

\begin{enumerate}[topsep=0pt]

\item Select the $RAE$ set (see Section~\ref{subsection:endpoints-selection}).

\item For each possible $RAE$ pair, measure the RTT on the direct path via single-packet pings.
We repeat this process 6 times as mentioned above, and calculate the median RTT per $RAE$ pair. In a time slot of 30 minutes, we send 6 consequent ping packets per pair with a time interval of 5 minutes.

\item Select a set of feasible relays per type. We apply the selection methodology of Section~\ref{section:colo-selection} for $COR$, \ref{section:planetlab-selection} for $PLR$, and Section~\ref{section:ripe-relay-selection} for $RAR$. To find only the feasible relays per $(RAE_1,RAE_2)$ pair we use the methodology of Section~\ref{subsection:limiting-relays}, based on the median RTTs on the $(RAE_1,RAE_2)$ paths calculated during Step (2). 

\item For each $(RAE_1,RAE_2)$ pair extracted from the $RAE$ set of
Step (1), and using the relays
from Step (3), we measure the RTT between $(RAE_1,RAE_2)$\footnote{The latency measurement for direct and relayed paths should be in sync. Thus, for each $(RAE_1,RAE_2)$ pair we recalculate the RTT on the corresponding direct path.}, $(RAE_1,Relay)$, 
and $(RAE_2,Relay)$
pairs, via single-packet pings.
We repeat this process 6 times, with a time interval of 5 minutes,
and calculate the median RTT per pair, based on at least 3 valid RTTs within a measurement window; thus allowing for meaningful median values. To infer the median RTT of a relayed path $(RAE_1,Relay,RAE_2)$ we stitch the associated median RTTs of $(RAE_1,Relay)$ and $(RAE_2,Relay)$.

\end{enumerate}

\section{Measurement Results} \label{section:measurement-results-analysis}

\begin{figure}[t!]
\centering
\includegraphics[scale=0.43]{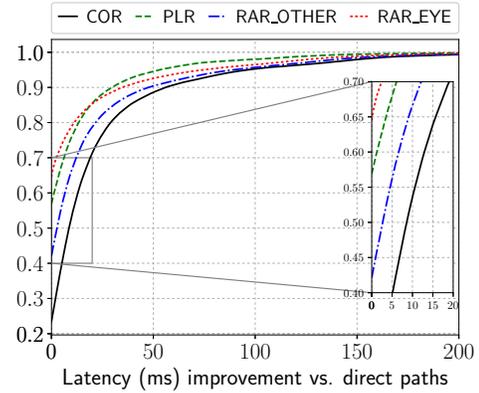}
\vspace{-3mm}
\caption{CDF of latency differences (RTT) \textit{vs.}~direct paths for the best relays
~(inducing minimal latency) per type per $RAE$ pair. Improvements between 1 and 200ms are shown (83\% of total cases). A few cases can reach up to 660ms.}
\label{fig:positive-full-cdf}
\vspace{-5mm}
\end{figure}

We ran the measurement workflow of Section~\ref{subsection:measurement-framework} for 45 rounds (20/4-17/5/2017), sending $\sim$8.7M~pings in total. We found $\sim$84\% of the destinations of the involved node pairs to be responsive with $\geq$3 ping replies per round. 
Next, we describe the most important insights related to the latency-wise performance 
of $\sim$29 million studied relayed paths vs.~$\sim$90K direct paths.

\textbf{Latency Improvements per Type.} Fig.~\ref{fig:positive-full-cdf} displays the CDF of the latency differences (in ms) of the best-performing
(i.e., inducing the least latency) relays per type, vs.~the direct paths over the entire set of measurements. We show the improved cases (83\% of total), where the relays yield lower-latency paths. We note that $COR$ paths perform better than direct in 76\% of the total cases, $RAR\_other$ in 58\%, $PLR$ in 43\% and $RAR\_eye$ in 35\%.
The latency improvements range from 1 to 200ms. A few outliers, such as communications involving very distant countries can witness even larger improvements\footnote{E.g., a path involving Colombia and Slovakia, observed reductions of 660ms when relayed through large European Colos.}.~Median improvements range between 12 and 14ms for all types. $COR$ and $RAR\_other$ yield improvements $>$100ms 
(which are critical for e.g., application service providers~\cite{amazonlatency}) in 6\% of the improved cases (5\% of total).
These gains stem solely from the discovery of fast TIV-enabled paths,
and do not consider other sources of latency that cut through the network stack~\cite{bozkurt2017internet}.
Note that $RAR\_eye$ and $PLR$ have very similar (low) performance, while $RAR\_eye$ and $RAR\_other$ differ significantly; the latter supports our intuition of differentiating between the two $RAR$ types. 
The difference between the best ($COR$) and the second-best overlay ($RAR\_other$) does not surpass 5-10ms, for the cases where both perform better than the direct paths. 
Therefore, the most important difference w.r.t.~performance is the percentage of cases where relays actually yield better-than-direct paths, while the improvement itself does not vary significantly across types. We further calculate a median of 8 $COR$, 3 $PLR$, 2 $RAR\_other$ and 2 $RAR\_eye$ relays that yield improvements for each $(RAE_1,RAE_2)$ pair, indicating a high redundancy of $COR$ relays.

\begin{figure}[t!]
\centering
\includegraphics[scale=0.43]{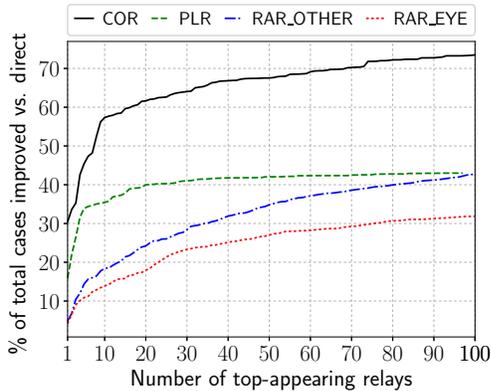}
\vspace{-3mm}
\caption{\% of total cases (pairwise communications) where relayed paths improve latency against direct paths, \textit{vs.} number of top relays (cut at top-100 relays 
for clarity).}
\vspace{-5mm}
\label{fig:relay-gain-vs-num}
\end{figure}

\begin{figure}[t!]
\centering
\includegraphics[scale=0.44]{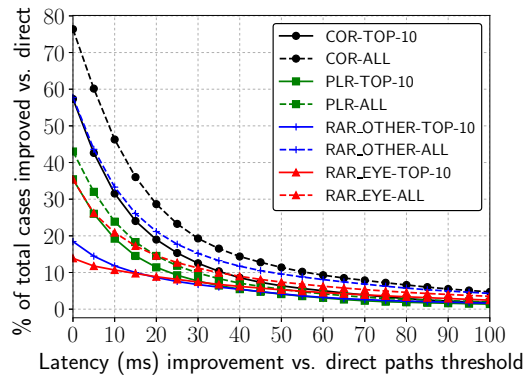}
\vspace{-3mm}
\caption{\% of total cases (pairwise communications) where relayed paths improve latency against direct paths (top-10/all relays), \textit{vs.} improvement threshold (cut at 100 ms). The best performance of each relay set is considered per case.}
\vspace{-5mm}
\label{fig:relay-gain-ms-top}
\end{figure}

\textbf{How Many Relays are Enough?} We next show what is the maximum benefit we can achieve per relay type, for a given number of relays. Fig.~\ref{fig:relay-gain-vs-num} shows the percentage of improved $(RAE_1,RAE_2)$ pairs (out of the total cases) vs.~the number of top relays (ranked according to their frequency of improvement) employed to achieve those improvements. 
The number of improved pairs increases rapidly with the first few $COR$ and $PLR$ relays; in particular, the top $COR$ relays are extremely beneficial (heavy hitters). In contrast, the number of improved pairs increases more smoothly for $RAR$ relays, which require $>>$100 relays 
to yield their top improvements (58\% of total cases, beyond x-axis bounds of Fig.~\ref{fig:relay-gain-vs-num}). Overall, $COR$ improve many more pairs with fewer relays. 
Specifically, 10 $COR$ relays (in 6 Colos) with latency improvement in $\sim$75\% of the improved cases (58\% of total) match the second-best performance (58\% of total cases for $RAR\_other$), which requires though $>>$100 $RAR\_other$. After the first 10 relays, the incremental benefit of additional $COR$ is decreasing fast. 
Fig.~\ref{fig:relay-gain-ms-top} shows the percentages of improved pairs (out of the total cases) vs.~the threshold of latency reduction that they surpass, when employing the top-10 and all relays of each type, respectively. The best performance of each relay set is considered per case. We see that the top-10 $COR$ perform better than the top-10 relays of all the other types, and follow closely the performance of all $RAR\_other$ relays (second-best performance after all $COR$). The gaps between top-10 and all relays differ per type; e.g., for $PLR$, this gap is minimal ($\sim$5\%), indicating very few well-performing relays among the relay set.
Interestingly, using only the top-10 $COR$ relays, $\sim$20\% of all pairs witness latency reductions larger than 20ms; the cost of this limited selection is only 30\% less --relative to all $COR$-- pairs surpassing 20ms (10\% being the exact percentage of ``missing'' total pairs).
We next examine the features of the top COR relays.

\textbf{Features of Top Facility Relays.}
Table~\ref{tab:top-fac} shows the facilities hosting the 20 top-appearing $COR$ relays, i.e., the ones with the highest frequency of improvement vs.~direct paths.
We augmented the facility data using information from PeeringDB~\cite{peeringdb}. Interestingly, we note that only 10 facilities actually contain the top-20 relays, and 4 of them are in the top-10 of PeeringDB w.r.t.~the number of colocated networks they host. All of them are colocated with at least 2 IXPs, with one facility
(Equinix Frankfurt) connected with 11 IXPs. At least 22 networks (ISPs, content/cloud providers, etc.) are colocated at each facility, with one (Telehouse North London) hosting 361 networks. All top facilities are either offering cloud services themselves, or are colocated with cloud providers; these data centers could be used e.g., to host VM-based relays~\cite{makkes, cai2016cronets}. In addition, the top facility relays reside in large metropolitan centers (and Internet hubs), mainly in Western Europe and North America. 

\textbf{Changing Countries and Paths.} Path inflation~\cite{spring2003causes}, induced by BGP policies, can lead to increased inter-domain latency between remote endpoints (residing, e.g., in different countries). We expect that this effect may prevent relays that are located close to the endpoints (e.g., same country) from using alternate, non-inflated paths. To verify this assumption, we compare relayed paths when the relay changes (and does not change) country w.r.t.~the endpoints;
we consider the min-latency relays per case. 
We observed that when the relay is in a different country than both endpoints\footnote{
Endpoints are located in different countries based on the selection of Section~\ref{subsection:endpoints-selection}.},
latency is lower than the direct path in 75\% of the cases for $COR$. In contrast, this number drops to 50\% if the relay is in the same country as one of the endpoints. Similar remarks apply for the other types, albeit with lower percentages. 
Also, out of the totally studied ($\sim$90K) $RAE$ pairs (and related direct paths), 74\% are inter-continental, indicating a set conducive to path inflation. Indeed, a significant fraction (19\%) of the total direct paths, turn up with RTTs that exceed 320ms (considered as threshold for poor VoIP performance~\cite{cisco,itu}). These values are in line with the work of Jiang \emph{et al.}~\cite{jiang2016via}. 
By employing only $COR$ relays, the fraction of paths over 320ms falls to 11\%.
 
\textbf{Stability over Time.} Regarding the evolution of the results \emph{in time}, we observed a consistent pattern, with $COR$ finding lower-latency paths in $>$75\% of the cases, $RAR\_other$ in $>$50\%, and $RAR\_eye$ and $PLR$ having a positive impact in less than 50\% of the cases in every measurement round. 
In general, in $\sim$50\% of the cases we found a 1-20ms improvement for $COR$.
The cumulative insights from Fig.~\ref{fig:positive-full-cdf} seem to apply consistently in time.
In fact, to further investigate the temporal stability of our observations,
we calculated the Coefficient of Variation (CV) for all the direct and relayed pairs for all measurements, as the standard deviation of the median RTTs of each pair divided by the pair's average of medians over time.
We observed that the CV ranges from 0\% to 40\%, and is less than 10\% in 90\% of the cases. 
This indicates stable, usable overlays.

\begin{table}[t!]
\centering
\caption{Facilities of top-20 Colo relays (ranked according to their frequency of presence in improved paths), and their location and connectivity characteristics.}
\vspace{-3mm}
\label{tab:top-fac}
\resizebox{\columnwidth}{!}{
\begin{tabular}{l|cccccc}
\hline
\multirow{2}{*}{\textbf{Facility Name (PDB ID)}} &{\textbf{\% of Improved}} &\multirow{2}{*}{\textbf{City (Country)}}
&\multirow{2}{*}{\textbf{\#Nets}} &\multirow{2}{*}{\textbf{\#IXPs}}  &{\textbf{Cloud}} &{\textbf{PDB}}\\ 
{} &{\textbf{Cases}} &{} &{} &{} &{\textbf{Services}} &{\textbf{top-10}}\\
\hline

{1) Telehouse North (34)} &{47} &{London (GB)} &{361} &{6} &{\cmark} &{\cmark} \\

{2) Equinix-AM7 (62)} &{46} &{Amsterdam (NL)} &{184} &{4} &{\cmark} &{\cmark} \\

{3) Nikhef (18)} &{34} &{Amsterdam (NL)} &{151} &{6} &{\cmark} &{\xmark} \\

{4) Equinix-FR5 (60)} &{30} &{Frankfurt (DE)} &{235} &{11} &{\cmark} &{\cmark} \\

{5) Telehouse West (835)} &{29} &{London (GB)} &{89} &{5} &{\cmark} &{\xmark} \\

{6) Digital Realty Telx (125)} &{29} &{Atlanta (US)} &{125} &{2} &{\cmark} &{\xmark} \\

{7) Incolocate (105)} &{29} &{Hamburg (DE)} &{22} &{3} &{\cmark} &{\xmark}\\

{8) Interxion (68)} &{27} &{Brussels (BE)} &{58} &{3} &{\cmark} &{\xmark}\\

{9) Digital Realty Telx (10)} &{22} &{New York (US)} &{112} &{5} &{\cmark} &{\xmark}\\

{10) Equinix-LD8 (45)} &{21} &{London (GB)} &{208} &{4} &{\cmark} &{\cmark}\\
\hline
\end{tabular}
}
\vspace{-4mm}
\end{table}

\section{Related Work} \label{section:related-work}

Researchers have tinkered with the idea of exploiting TIVs to improve inter-domain routing for the last two decades.
After the early pioneers with the Detour Framework~\cite{savage1999detour,savage1999end},
Andersen \emph{et al.}~\cite{andersen} introduce Resilient Overlay Network (RON) to form resilient and---potentially faster---paths compared to the default BGP paths. Similar insights remain timely~\cite{haq2017measuring}, albeit by exploiting inter-continental cloud-terminated paths.

VIA~\cite{jiang2016via} aims at improving Internet telephony by employing classic overlay techniques to relay calls. They show that an oracle-based overlay can potentially improve up to 53\% of calls whose quality is impacted by poor network performance.
Their relay selection strategy uses call history information, and is based on the empirical observation that even though a prediction-based approach may not identify the optimal relay, it is likely to exist in the top few predicted relays.

Regarding the number of relays per relayed path, useful insights  are provided by Han \emph{et al.}~\cite{han} and Le \emph{et al.}~\cite{le}.
Both works support our approach to consider only 1-relay paths as adequate to reduce latency, compared to $N$-relay paths ($N\geq2$).

ARROW \cite{peter}, an inter-domain routing approach based on waypoints (i.e., relay routers within ISPs), allows users to 
set up reliable and secure e2e tunnels.
While for about 20\% of the ARROW cases, the e2e latency increases (up to 20\%), the performance of most paths may actually be improved when routed via ARROW waypoints. 
On the other hand, Lumezanu \emph{et al.}~\cite{lumezanu2,lumezanu1} analyze TIVs, concluding that even though faster inter-domain paths exist, their utilization can be prevented by business drivers of the ISPs themselves.

MeTRO~\cite{makkes} aims to offer QoS between endpoints, using virtual routers hosted in Amazon EC2~\cite{ec2} and Bright Box~\cite{brightbox} data centers as cloud relays. Latency improvements exist for 58\% of the cases, while the best performing relays are close to large IXPs.
In contrast to our work, no extensive comparison of overlay positioning is performed, to understand the location impact on the relay selection strategy. Similarly to MeTRO, Cai \emph{et al.}~\cite{cai2016cronets} propose cloud-routed overlay networks (CRONets), to maximize throughput.
Results show that CRONets consistently help for 78\% of the cases.
While MeTRO and CRONets relays are cloud-hosted,  
one of our goals is to suggest a large-scale methodology for measuring inter-domain paths passing through \emph{diverse} relay types.
To this end, we exploit pingable IP addresses of interfaces located in Colos, and we concatenate the latency of individual hops (endpoint to relay to endpoint). 
Since these interfaces do not have to be under our administrative control, they are not associated with any costs, therefore our methodology can scale seamlessly. 

In summary, we identify a tendency towards inter-domain overlay networks, using relays in data centers~\cite{cai2016cronets,jiang2016via,le,makkes}, ISPs~\cite{han,peter},
or at the last mile~\cite{andersen,haq2017measuring}.
By exploiting TIVs~\cite{lumezanu2,lumezanu1} to reduce inter-domain latency,  results show an improvement of latency metrics when overlay paths are employed, as compared to direct BGP-based paths. 
It is worth mentioning that the use of overlays requires a delicate balance between overlay-based optimization and policy-driven TE (e.g., on the enterprise level~\cite{lee2008improving}), to avoid potential policy conflicts~\cite{koutsoupias2009worst, qiu2003selfish, liu2005interaction} with monetary impact. However, our work focuses on strategically constructing and evaluating relayed paths for end-users and application providers; in particular, employing relays at Colos, not explored in previous works.

\section{Conclusions \& Future Work} \label{section:conclusions-future}

The Internet is changing. Requirements for low-latency video distribution have driven the flattening of the Internet topology in recent years, resulting in short distances and dense fabrics at interconnection facilities~\cite{labovitz}. This effect coupled with the emergence of numerous cloud providers, residing at Colos, is opening up the largest Colos of the Internet to end-users and application service providers, who can now easily host their services there. In this paper, we ask how this \emph{democratization} of large Colos affects latency for services that use relays. We performed an Internet-wide measurement study, spanning 1 month, employing different types of relays which serve endpoints located at the last mile. 
We showed that Colos are useful locations to host relays, taking advantage of their high connectivity and core locations to discover low-latency TIV paths that are faster than the direct ones. A few Colo-based relays are found to improve many more ($>$20\%) of the studied cases than one order of magnitude more PlanetLab or eyeball relays.

\textbf{Future Work.} We plan to investigate the following:

(i) The key factor(s) due to which Colos perform so well as relays. Even though a preliminary analysis has already been conducted in this work, the exact root-causes remain subject to further research.

(ii) The underlying reasons for the relatively good performance of $RAR\_other$ relays. RIPE Atlas is known to have a significant deployment even in commercial (core) networks. We plan to further examine the networks where these nodes are present.

(iii) Regional effects uncovered via traceroute measurements. For example, we intend to investigate potential correlations between the characteristics of the countries traversed by relayed paths and the achieved latency, as well as between the latency and the proximity of endpoints/relays to submarine cable landing points~\cite{submarine}.

\textbf{Software and Datasets.} The software used to run, analyze and visualize the measurements presented in this paper is publicly available, together with the collected measurement data~\cite{kotronis2017shortcuts}. 

\textbf{Acknowledgements.} 
This work has been funded by the \grantsponsor{ERCNetVolution}{EU Research Council}{\url{https://erc.europa.eu/}} Grant Agreement no. \grantnum{ERCNetVolution}{338402}.

\balance
\bibliographystyle{acm}
\setlength{\itemsep}{0pt}
\bibliography{biblio}

\end{document}